\documentclass[12pt]{article}
\usepackage{epsf,amssymb,latexsym,enumerate,cite,shadow,array,color}
\usepackage{graphicx,wasysym}

\usepackage[pdftex]{hyperref}
\usepackage{url}

\newcommand{\be}{\begin{equation}}
\newcommand{\ee}{\end{equation}}

\begin{document}

\begin{titlepage}

\hskip 1cm

\vskip 3cm

\begin{center}
{\LARGE \textbf{A brief history of the multiverse}}

\vskip 1.5cm

\

{\bf  Andrei Linde} 
\vskip 0.5cm
{\small\sl\noindent SITP and Department of Physics, Stanford University, \\
Stanford, CA
94305 USA\\
}
\end{center}
\vskip 1.5 cm

\

\begin{abstract}

The theory of the inflationary multiverse changes the way we think about our place in the world. According to its most popular version, our world may consist of infinitely many exponentially large parts, exhibiting different sets of low-energy laws of physics. Since these parts are extremely large, the interior of each of them behaves as if it were a separate universe, practically unaffected by the rest of the world. This picture, combined with the theory of eternal inflation and anthropic considerations, may help   to solve many difficult problems of modern physics, including the cosmological constant problem. In this article I will briefly describe this theory and provide links to the some hard to find papers written during the first few years of the development of the inflationary multiverse scenario. 
  
 \end{abstract}

\vspace{24pt}
\end{titlepage}

 \tableofcontents{}
\parskip 5.5pt
\hskip -0.5cm {\bf 7. Addendum: Excerpts from some early papers \hskip 2cm ~~~~~~~~~~~~~~~~~~~~~~~~~~~19}

\parskip 5pt

\section{Introduction}

The theory of inflationary multiverse is based on unification of inflationary cosmology, anthropic considerations, and particle physics. Its most advanced versions include a combination of eternal inflation and string theory into what is now called ``string theory landscape.''
This theory is still `work in progress,'  and the pendulum of public opinion with respect to it swings with a very large amplitude. Some people love this theory,   others hate it and write papers defending the integrity of physics.   It does not help much that the word ``multiverse'' is  used differently by different people.
In this situation it may be useful to remember what exactly we are talking about and why   this theory was invented. 

This is not an easy task. Until the mid-90's the cosmological anthropic principle pioneered by Dicke, Carter, Rees, Barrow, Rozental and others remained very unpopular. For example, we know that the proton mass is almost exactly equal to the neutron mass. If the proton were 1\% heavier or lighter, life as we know it would be impossible. Similarly, one cannot significantly change the electron charge and its mass without making the universe unsuitable for life.  
But if physical parameters  are just constants in the Lagrangian, nothing can change them. Therefore the standard lore was that one should avoid using anthropic arguments for explaining fundamental properties of our world.

As a result, some of the key ideas relating to each other inflation and anthropic considerations originally were expressed in a rather cryptic form and scattered among preprints, conference proceedings and  old journals which are hard to find.  In this paper I will briefly describe the history of the evolution of these ideas and provide links to  scanned versions of some of the original publications which are especially difficult to find. For convenience, I will also provide TeX versions of some parts of these papers in the Addendum to this paper.

Historically, there were many different versions of the theory of the multiverse based on the many-world interpretation of quantum mechanics \cite{Everett:1957hd} and quantum cosmology \cite{DeWitt:1967yk},  on the theory of creation of the universe `from nothing' \cite{Vilenkin:1982de}, and on the investigation of the Hartle-Hawking wave function \cite{Hartle:1983ai}. These ideas are most powerful, but their consistent implementation requires deep understanding of difficult conceptual issues of quantum cosmology.  Moreover, quantum cosmology by itself does not allow us to change fundamental constants.  

Therefore the main progress in the development of the theory of the inflationary multiverse was achieved in a different, conceptually simpler context. 
To explain it, let us remember that one of the starting points of the pre-inflationary cosmology was that the universe is globally uniform. This was the so-called `cosmological principle', which was invoked  by many people, from Newton to Einstein, to account for the observed large-scale homogeneity of the universe. 
The physical mechanism explaining the homogeneity of our part of the world was provided by inflationary theory. Surprisingly enough, this theory made the cosmological principle obsolete.

 Indeed, the main idea of inflationary cosmology is to make our part of the universe homogeneous by stretching any pre-existing inhomogeneities and by placing all possible `defects,' such as domain walls and monopoles, far away from us, thus rendering them unobservable. If the universe consists of different parts, each of these parts after inflation may become locally homogeneous and so large that its inhabitants will not see other parts of the universe, so they may conclude, incorrectly, that the universe looks the same everywhere. However, properties of different parts of the universe may be dramatically different. In this sense, the universe effectively becomes a multiverse consisting of different exponentially large locally homogeneous  parts  with different properties. 
To distinguish these exponentially large parts of our world from more speculative `other universes' entirely disconnected from each other, I called these parts `mini-universes,'  others call them `pocket universes'. Eventually, we started using the word `multiverse', or `inflationary multiverse' to describe the world consisting of many different `mini-universes', or `pocket universes'.

 An advanced version of 
this scenario describes our world as an eternally growing self-reproducing fractal consisting of many locally homogeneous parts (mini-universes). If the fundamental theory of all interactions has many different vacuum states, or allows different types of compactification, the laws of the low-energy physics and even the dimensionality of space in each of these mini-universes may be different.  This provided, for the first time, a simple scientific interpretation of the anthropic principle,  which did not rely on the possible existence of `other universes': We can live only in those parts of the world  which can support life as we know it, so there is a correlation between our own properties, and the properties of the part of the world that we can observe. 

\section{Inflationary Multiverse: The first models}

This combination of eternal inflation  and the anthropic principle, which we will concentrate upon in this paper, was first proposed in 1982 in my preprint written during the famous Nuffield Symposium on inflationary cosmology \cite{Linde:1982ur}. 
It was argued there that  an eternally inflating universe ``contains an infinite number of mini-universes (bubbles) of different size, and in each of these universes the masses of particles, coupling constants, etc. may be different due to the possibility of different symmetry breaking patterns inside different bubbles. This may give us a
possible basis for some kind of Weak Anthropic Principle: There is an
infinite number of causally unconnected mini-universes inside our universe,
and life exists only in sufficiently suitable ones.''  A full text of the preprint can be found at \url{http://www.stanford.edu/~alinde/1982.pdf} and also in the Addendum to the present paper. 

This idea, which plays the central role in the theory of inflationary multiverse, was further extended in my contribution to the proceedings of  the Nuffield  Symposium \cite{Linde:1982gg}. In addition to the discussion of various types of symmetry breaking in different parts of the multiverse,  I also revisited there an old observation by Paul Ehrenfest   that life as we know it may exist only in a four-dimensional space-time, because planetary and atomic systems would be unstable for $d\not = 4$ \cite{Ehresnfest}. I argued in \cite{Linde:1982gg} that this idea finally acquires a well defined physical meaning in inflationary theory combined with the idea of  spontaneous compactification: ``In the context of this scenario it would be sufficient that the compactification to the space $d = 4$ 
is {\it possible}, but there is no need for the four-dimensional space to be {\it the only} possible space after the compactification. Indeed, if the compactification to the space $d = 4$ is possible, there will be infinitely many 
mini-universes with $d = 4$ in which intelligent life can exist.'' 

Being impressed by free lunches provided to the participants of the Nuffield Symposium, I summarized the main consequence  of this scenario as follows: ``As was claimed by Guth (1982), the inflationary universe is the only example  of a free lunch (all matter in this scenario is created from the unstable vacuum). Now we can add that the inflationary universe is the only lunch at which all possible dishes are available''  \cite{Linde:1982gg}, see the Addendum and  \url{http://www.stanford.edu/~alinde/LindeNuffield.pdf}.

Ref. \cite{Linde:1982ur} described  eternal inflation in a particular version of the new inflationary scenario \cite{Linde:1981mu} where the inflaton potential has a shallow minimum at the top. The regime of eternal inflation occurs in the old inflationary theory as well, but there it was considered a major obstacle precluding a consistent realization of inflationary cosmology \cite{Guth:1980zm,Guth:1982pn}. A possible existence of this regime in  new inflation   was first mentioned by Steinhardt in his talk at the Nuffield symposium. A brief discussion of this idea, as well as of my paper \cite{Linde:1982ur}, is contained in his contribution to the Proceedings of the Nuffield Symposium \cite{Steinhardt:1982kg}. However, he did not discuss  the use of the anthropic principle; in fact he opposes the theory of inflationary multiverse in strongest possible terms \cite{Steinhardt:2011zza}.
A more general 
approach to the theory of eternal inflation  in the new inflationary scenario was developed in 1983 by Vilenkin \cite{Vilenkin:1983xq}. He explained that inflationary quantum fluctuations may keep the field at the top of the potential in some parts of the universe even if the potential does not have a local minimum at the top.  

The first practical application of the anthropic principle in the context of inflationary cosmology was given  in \cite{Linde:1984je}.   It was shown there that inflationary fluctuations can induce transitions between different vacua in supersymmetric grand unified theories. These transitions divide the universe into exponentially large parts with different types of symmetry breaking, and therefore with different laws of low-energy physics. If not for this mechanism,  the hot universe would stay forever  in the SU(5) symmetric vacuum, unsuitable for  life as we know it.

None of these developments attracted much attention at that time, in part because the anthropic principle was very unpopular, and in part because the new inflationary scenario did not work and was replaced by the chaotic inflation scenario \cite{Linde:1983gd}. The simplest versions of the chaotic inflation scenario describe inflation far away from any extrema of the inflaton potential, so the previous results obtained in \cite{Steinhardt:1982kg,Linde:1982ur,Vilenkin:1983xq} did not apply to it. In addition, the energy scale of  eternal inflation in the new inflation scenario was extremely small as compared to the Planck scale. Therefore it was not clear  whether inflationary fluctuations could be powerful enough to probe the inner structure of space and divide the universe into different parts with different types of compactification, as proposed in  \cite{Linde:1982gg}. This problem was solved in 1986  with the discovery of the regime of eternal inflation in the chaotic inflation scenario \cite{Linde:1986fc,Linde:1986fd}.

\section{Eternal chaotic inflation}


This proposal was initiated by the discovery of the regime of eternal inflation in the chaotic inflation scenario   \cite{Linde:1986fc,Linde:1986fd}. (In fact, the name ``eternal inflation'' was first introduced in these two papers.) It was shown there that eternal inflation is a generic regime which is possible in any theory with a sufficiently flat potential. This included not only new inflation, but any chaotic inflation model with polynomial potentials. More generally, eternal inflation occurs in any inflationary regime where the amplitude of density perturbations may become $O(1)$ for some values of the fields. The theory of this effect is briefly described in Appendix A of this paper.

One of the most important features of this regime was the possibility of eternal inflation at arbitrary values of energy density, up to the Planck energy density. This implied that inflationary perturbations of all other fields produced at that epoch could be large enough to bring us from one vacuum state to another, and even from one type of compactification to another, practically irrespectively of the size of the barrier separating different vacua.  
This realization was one of the strongest emotional shocks of my life, which forced me to delay and completely re-write my book on inflationary cosmology, which I was preparing at that time \cite{Linde:2005ht}. 

First of all, the universe in this scenario becomes immortal. Not only that, but in the eternal process of self-reproduction it re-creates itself in all possible ways, probing all laws of low-energy physics compatible with the underlying physical theory. This opens the possibility  to implement the anthropic principle in its strongest form, using the analogy with mutations of the laws of physics and the Darwinian approach  \cite{Linde:1987aa}. The main idea was illustrated in   \cite{Linde:1987aa} by an image of a multicolored self-reproducing universe, which  gradually became the standard symbol of the inflationary multiverse. Fig. 1 shows its version published in \cite{Linde:1987aa}, with the original figure caption. By looking at this figure, one can instantly realize that  eternal inflation transforms  the universe into a huge eternally growing fractal. These are not just words: In 1987 Aryal and Vilenkin calculated fractal dimension of the universe in the new inflation scenario \cite{Aryal:1987vn}, and then, few years later, a similar calculation was performed for the eternal chaotic inflation \cite{Linde:1993nz}.

\begin{figure}[h!]
\centering
{\hspace{5mm}
\includegraphics[scale=.3]{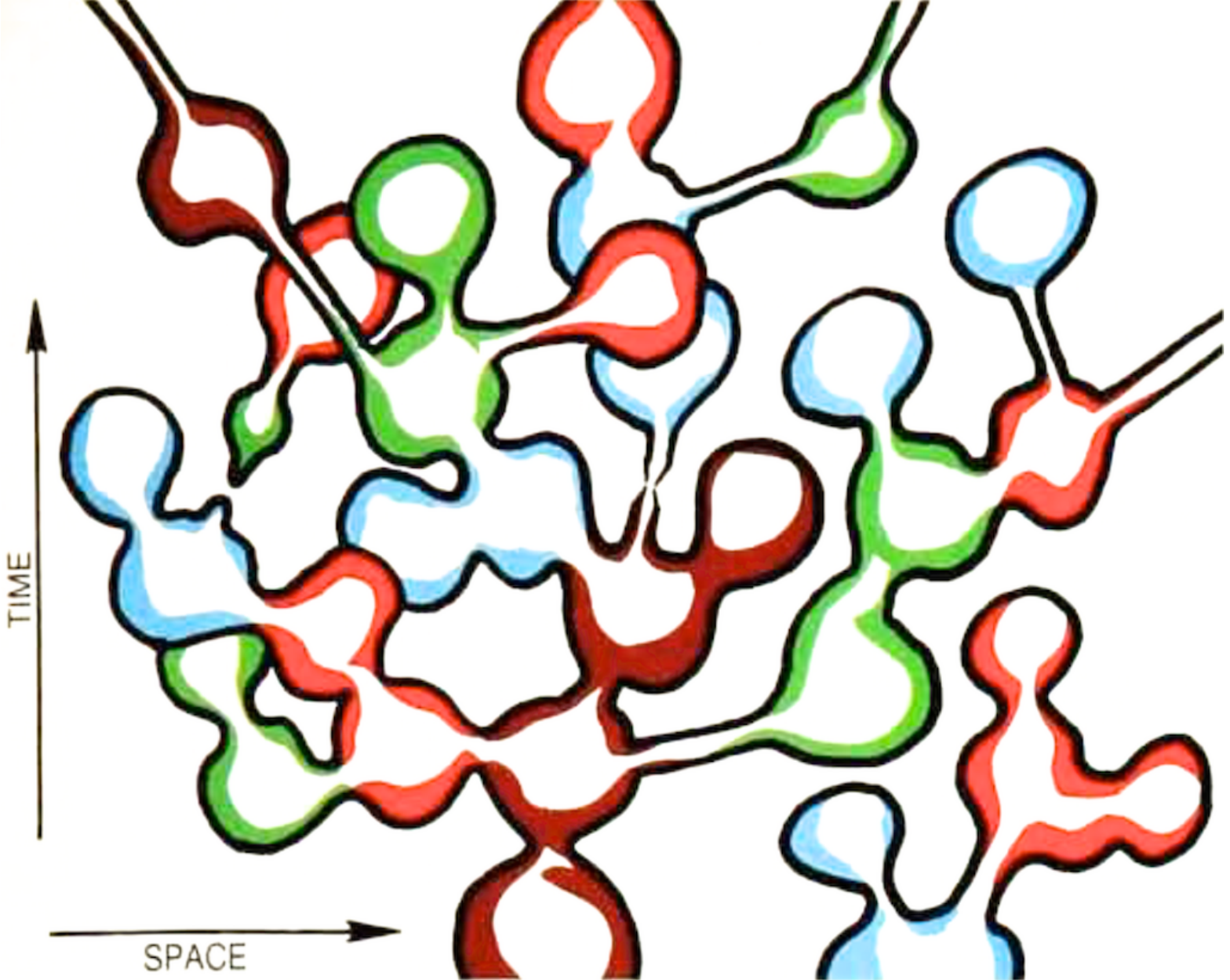}
\label{figT}}~~~~~
\caption{\footnotesize Global structure of a chaotic, self-reproducing inflationary universe. Locally (out to the $10^{10}$ light-year horizon) the universe looks quite homogeneous, but its global structure is complex. Mini-universes at the Planck energy density are ``mutants" that may forget completely the ``genetic code" (color) of their parent universe. They may even have a
different space-time dimensionality. The typical thickness of a tube connecting two mini- universes after inflation is exponentially large, but if it corresponds to a compactified inflationary universe it can be as thin as the Planck length ($\sim 10^{-33}$ cm). If the tube then evaporates by Hawking radiation, the parent and offspring mini-universes have lost their umbilical space-time connection. (Reproduced from my paper  ``Particle Physics and Inflationary Cosmology,''
  Physics Today {\bf 40}, 61 (1987), with the permission of the American Institute of Physics.)}
\end{figure}

I anticipated that the most interesting consequences of this scenario will be achieved in string theory and suggested that ``an enormously large number of possible types of compactification which exist e.g. in the theories of superstrings should be considered not as a difficulty but as a virtue of these theories, since it increases the probability of the existence of mini-universes in which life of our type may appear''  \cite{Linde:1986fd}, see the Addendum to the present paper.
But time was not yet ripe for eternal inflation in string theory because we did not know   how to stabilize multiple string theory vacua  discussed e.g. in \cite{Lerche:1986cx}. 
The emphasis shifted to other aspects of the inflationary multiverse scenario. In particular, in order to make the anthropic selection viable, it was necessary to find physical mechanisms that would allow   physical parameters to take  different values in the context of the same physical theory. And here inflation helped to reveal many new possibilities.

For example, it was found that in certain mechanisms of baryogenesis, the ratio $n_{b}/n_{\gamma}$ was controlled by a light scalar field which experienced quantum fluctuations during inflation  \cite{Affleck:1984fy}. As a result, the parameter $n_{b}/n_{\gamma}$ in this scenario may take different values in different parts of the universe  \cite{Linde:1985gh}. But why do we live in the part of the universe where $n_{b}/n_{\gamma}$ is extremely small? A possible answer  was that the large scale structure of the universe in its parts with  relatively large ratio $n_{b}/n_{\gamma}$ would be very different, which would make emergence of life as we know it much less likely. 
A similar argument was used in \cite{Linde:1987bx} to remove the standard constraints on the axion mass and explain the large ratio of dark matter to normal matter.  Recently these ideas were revived and developed by several authors, see e.g. \cite{Tegmark:2005dy,Freivogel:2008qc,Bousso:2013rda}. Investigation of inflationary perturbations in the Brans-Dicke theory has shown that the effective gravitational constant and the amplitude of perturbations of metric can take different values in different part of the post-inflationary universe \cite{Linde:1989tz}.  Ref. \cite{Linde:1988yp} contained an explicit realization of the possibility that our world may consist of different parts with different number of large (uncompactified) dimensions. Thus it was realized that many features of our world may indeed have anthropic origin. 
Most of these results were summarized in 1990 in my books ``Particle physics and inflationary cosmology'' \cite{Linde:2005ht} and ``Inflation and quantum cosmology'' \cite{book2}. 

The first ten years of the journey through the multiverse were very exciting but rather lonely. Not many people were interested in this theory; almost all of my papers on inflationary multiverse have been written without any collaborators. The editor of my book  \cite{Linde:2005ht} recommended me to delete the part on the anthropic principle, saying that otherwise I will lose respect of my colleagues. I replied that if I do it, I will lose self-respect, and eventually she agreed to keep it. It appears though that the risk was very minimal anyway, since this part was in the very end of the book, in the chapter on quantum cosmology, which was of little interest to most of the readers at that time.

In the beginning of the 90's the situation began to change. 
In 1993  a detailed theory describing two different probability measures in the eternally inflating universe was proposed in a series of my papers written in collaboration with Arthur Mezhlumian, Juan Garcia-Bellido and Dimitri Linde \cite{Linde:1993nz,GarciaBellido:1993wn}.  In 1994 Alex Vilenkin resumed his investigation of eternal inflation. His prolific work, as well as the work of his numerous talented collaborators, invigorated the theory of inflationary multiverse, bringing to it new interesting choices of the probability measure and anthropic considerations, see e.g. \cite{Vilenkin:1994ua} and his book ``Many worlds in one'' \cite{Vilbook}. But despite  this exciting progress, the theory of inflationary multiverse was developed only by a small group of experts and remained of no interest for the general physical community. And the crucial part of this theory, the anthropic considerations, was almost universally despised. 
The situation changed dramatically only after the discovery of the cosmological constant/dark energy and the development of the string theory landscape.

\section{Multiverse, string theory landscape, and the cosmological constant problem}

The  discovery of dark energy in 1998 \cite{Riess:1998cb,Perlmutter:1998np} pushed the cosmological constant problem to the forefront of research. The observers found that empty space is not entirely empty, it has  tiny energy density $\sim 10^{{-29}}$ g/cm$^{3}$. This minuscule number is 120 orders of magnitude smaller than the Planck density, it is  40 orders of magnitude  smaller than nuclear density and 29 orders of magnitude smaller than density of water.  In the early universe, the energy density of usual matter was many orders of magnitude greater than $\sim 10^{{-29}}$ g/cm$^{3}$; in the distant future it will be much smaller than the vacuum energy density.  But for some mysterious reason it is of the same order of magnitude than the energy density of matter of the universe at present time. This discovery triggered an unexpected chain of events in theoretical physics.

For many decades theorists were unsuccessfully trying to find a theory which would explain why the cosmological constant, representing the vacuum energy density, is exactly zero. But we could not do it; it was a spectacular failure.  After 1998, we faced a much more complicated problem: It was necessary to explain why vacuum energy/cosmological constant is not exactly zero but is extremely small, 
and why  this constant  is of the same order of magnitude as the density of normal matter at the present epoch. The only known way to address these two problems without invoking incredible fine-tuning was related to the anthropic principle, and, therefore, to the theory of the multiverse.

The possibility that the value of the vacuum energy may be determined by the anthropic considerations was known  long ago.  For example, back in 1981 Davies and Unwin \cite{DaviesUnwin} noted: { ``In the absence of a fundamental reason why $\Lambda$ should be so small, a possible anthropic explanation suggests itself. Perhaps the excessive smallness of $\Lambda$ is a feature that only characterizes our particular region of the universe. In other regions this fine-tuning fails and $\Lambda$ assumes much greater values. But in such regions $\Lambda$ would dominate the gravitational dynamics, leading to exponential expansion, or (for negative values) collapse into anti-de Sitter space. Probably, values that differ by more than a few orders of magnitude from the observed upper limit in our region would be sufficient to prevent the formation of galaxies, and hence organic life.''} 
Similar arguments were given in \cite{Linde:1984ir,Sakharov:1984ir,Banks:1984cw,BarrowTipler,Linde:1986dq}. But the real derivation of the  anthropic bound on the positive cosmological constant was given for the first time by Weinberg in his famous paper of 1987 \cite{Weinberg:1987dv}, and it was further strengthened in  \cite{Martel:1997vi} and several subsequent publications. 

The remaining problem was to find a physical mechanism that would allow the cosmological constant to vary and to become extraordinarily  small. Even the first part of this problem was rather difficult because the cosmological constant was considered by many to be just a constant in the gravitational action, so it was not quite clear whether it makes any sense to consider universes with different values of $\Lambda$. Then in 1974 it was realized that a constant scalar field may play the role of a cosmological constant, taking different values in different places at different time \cite{Linde:1974at}. But having 10 or 100 different vacua with different values of  $\Lambda$ would not explain why it is as small as $10^{{-120}}$ in Planck units.
To address this issue, Davies and Unwin  \cite{DaviesUnwin}  suggested to use a twisted configuration of the scalar field interpolating between two minima of the Higgs-type potential, in a hope to represent  different values of the effective cosmological constant by the slowly changing spatial distribution of a scalar field. However, for the parameters used in \cite{DaviesUnwin} this distribution represented a narrow domain wall, which did not lead to the cancellation of $\Lambda$. In an attempt to explain the smallness of the cosmological constant, Banks introduced one of the first models of dark energy, but after studying it he concluded  \cite{Banks:1984cw}: ``This is a deadblow for the present model. Even the anthropic principle cannot save it.'' 
In what follows I will describe three different ideas proposed in 1984-1986  \cite{Linde:1984ir,Sakharov:1984ir,Linde:1986dq}, which found their way to modern discussions of the cosmological constant problem.

1) The first of these proposals \cite{Linde:1984ir}  was based on quantum cosmology and the possibility of creation of the universe ``from nothing''. Initially, papers on quantum creation of the universe  \cite{Vilenkin:1982de,Hartle:1983ai} suggested that the probability of creation of a closed inflationary universe with the inflaton energy density $V(\phi)$ is given by $e^{24\pi^{2}/V(\phi)}$. This implied that the probability of quantum creation of an inflationary universe is exponentially small, suggesting that it is much easier to create a huge empty universe of a present size, with a tiny value of the cosmological constant. However, the subsequent investigation indicated that the probability of the universe creation is in fact given by  $e^{-24\pi^{2}/V(\phi)}$    \cite{Linde:1983mx,Linde:1984ir,Vilenkin:1984wp}. 
This provided an ideal framework for realization of initial conditions for chaotic inflation, where $V$ can be large \cite{Linde:1983mx,Linde:1984ir}. There could be a problem for those models where inflation is possible only for $V \ll 1$ and the term $e^{-24\pi^{2}/V(\phi)}$ is exponentially small, but one could alleviate this problem using  anthropic considerations \cite{Vilenkin:1984wp}.  Moreover, an investigation of quantum creation of compact open or flat universes with nontrivial topology indicated that this process is not exponentially suppressed even for $V \ll 1$  \cite{Zeldovich:1984vk,Coule:1999wg,Linde:2004nz}. 
And once inflation starts, its nontrivial topology becomes practically unobservable: The universe becomes locally uniform and isotropic.

In order to use these results for solving  the cosmological constant problem, one may  consider a combined contribution of  scalar fields $\phi$ and fluxes (antisymmetric tensor fields $F_{{\mu\nu\lambda\sigma}}$) to vacuum energy. At the classical level the fields $F_{{\mu\nu\lambda\sigma}}$ take constant values all over the universe and contribute to the cosmological constant \cite{Hawking:1984hk}.
At that time it was not known that these fields may change their values due to tunneling with bubble formation \cite{Brown:1987dd}. Therefore one could expect that the cosmological constant is just a constant which cannot  change. However, in quantum cosmology this is not necessarily an obstacle. One may study the probability of creation of different universes with different values of $\phi$ and $F_{{\mu\nu\lambda\sigma}}$. If one uses the expression for the probability $e^{24\pi^{2}/V}$ \cite{Vilenkin:1982de,Hartle:1983ai}, one could conclude that $\Lambda$ must vanish  \cite{Hawking:1984hk}, which is not the case according to the observational data. However, if one uses the expression $e^{-24\pi^{2}/V}$  \cite{Linde:1983mx,Linde:1984ir,Vilenkin:1984wp}, where $V = V(\phi) +V(F)$, one finds that the probability to live in a universe with a given value of $\Lambda$ has a flat distribution as a function of $\Lambda$, which is exactly what we need for the anthropic solution of the cosmological constant problem, see the very end of the paper   \cite{Linde:1984ir}, and the Addendum to the present paper.

2) The second mechanism was proposed by Sakharov \cite{Sakharov:1984ir}. He mentioned, following   \cite{Linde:1982gg}, that   the universe may consist of many different parts with different types of compactification. Then he argued that if the number of compactified dimensions is sufficiently large, the number of different types of compactifications can be exponentially large. He emphasized that if this number is large enough, the typical energy gap between different levels can be extremely small, which may allow to explain the smallness of the cosmological constant by using anthropic considerations, see  Section 4 in \url{http://www.stanford.edu/~alinde/Sakharov1984.pdf} and the Addendum.

3) The third mechanism was proposed in \cite{Linde:1986dq}. It was based on a combination of eternal inflation driven by the inflaton field $\phi$ and a subsequent slow roll of what was later called `quintessence' field $\Phi$. The role of eternal inflation was to generate perturbations of the field $\Phi$, which then give this field different values in different exponentially large parts of the universe. As a result, the universe becomes divided into different parts with a flat probability distribution for different values of the effective cosmological constant. Once again, this provided a possibility to use anthropic considerations for solving the cosmological constant problem, see  \url{http://www.stanford.edu/~alinde/1986300yrsgrav.pdf}  and the Addendum.

All of these ideas did not attract much interest until the discovery of the cosmological constant in 1998. This discovery created a lot of tension, especially since the cosmological constant was positive, and at that time we did not know any way of describing a positive cosmological constant in string theory.
 
 An important step was made in 2000 by Bousso and Polchinski \cite{Bousso:2000xa}, who  proposed a string theory motivated model explaining a possible reason for an exponentially large number of vacua with different values of the cosmological constant. This could provide an anthropic solution to the cosmological constant problem. The idea was brilliant, preserving some  elements of the  suggestions made in \cite{Linde:1984ir,Sakharov:1984ir},  but going much further. But at that time we still did not know how one could consistently construct any stable or metastable string theory vacua with positive vacuum energy (de Sitter vacua). 
 
 Eventually, several interesting mechanisms of vacuum stabilization in string theory have been proposed, see in particular \cite{Str}. The most developed one, the so-called KKLT construction  \cite{Kachru:2003aw}, was proposed  in 2003. We found a possible way to stabilize string theory vacua, but we also found that all de Sitter vacua in string theory are not absolutely stable but metastable, and all barriers separating these vacua can be penetrated with finite probability.  This was the crucial observation, which implied that the universe can exponentially expand in any of these metastable de Sitter states as in the eternal inflation scenario, and tunnel from each of them to any other string theory vacuum. I will describe the proof of this important statement in the Appendix B.  Estimates made in  \cite{Douglas} have shown that the total number of such vacua can be as large as  $10^{500}$. Even if one starts with the universe in one of these vacua,  a combination of eternal inflation and the inevitable tunneling between these vacua  should create a multiverse consisting of parts where all of these de Sitter vacua are represented. Lenny Susskind called this scenario `the string theory landscape'  \cite{Susskind:2003kw}.    An incredible richness of this landscape may help us to solve the cosmological constant problem, as well as many other problems which required anthropic explanation. This perfectly matched  the earlier expectations  expressed  in \cite{Linde:1986fd}.

After these developments, the general attitude towards inflationary multiverse and anthropic considerations changed.
Powerful efforts by  Vilenkin,  Susskind,  Bousso, Guth, Shenker,  Hall,  Nomura and many others rapidly transformed this field into a vibrant and rapidly developing branch of theoretical physics. String theory underpinnings of  vacuum stabilization, uplifting, inflation, and other aspects of the multiverse scenario received strong boost from subsequent works of  Silverstein, Kachru, Polchinski, Kallosh and others; see especially the recent series of papers \cite{Kallosh:2014wsa,Bergshoeff:2015jxa,Michel:2014lva,Kallosh:2015nia,Polchinski:2015bea}. 
 There is a  broad agreement between different scientists about the general structure of the inflationary multiverse, and there are even some attempts to expand the concept of the multiverse from physics to mathematics \cite{Tegmark}.  But much more work is required if we want to use our new knowledge for making unambiguous scientific predictions and explaining our observations. Indeed, if eternal inflation produces infinitely many red universes and infinitely many green universes, then which of these infinite sets of universes  will dominate? This is called the measure problem \cite{Linde:1993nz}. 
 
This 
problem is not specific to the multiverse, it is just a byproduct of dealing with infinities. 
 Consider for example the usual flat or open Friedmann universe. It is  infinite, and therefore it contains infinite number of planets, infinite number of cities, and infinite number of rooms. Since there are infinitely many of them, there are also infinitely many ``bad'' rooms where all oxygen molecules simultaneously move into a corner of the room, and everybody suffocates. The probability of this process in each particular room is incredibly small, but if one takes into account that the number of ``bad'' rooms is infinite, the statements about the probabilities become ambiguous. For example, one can take one ``bad'' room and one ``normal'' room, then yet another ``bad'' and ``normal'' room,  continue this counting for indefinitely long time and conclude that  50\% of all rooms are ``bad.'' 

This is one of the many paradoxes which appear when one compares infinities. One  way to deal with it is to follow the lead of quantum mechanics and make only those predictions that an observer can actually verify, under initial conditions that he/she should know/prepare before making predictions. This makes the subset of accessible rooms finite, and   the normal expectations based on the standard laws of thermodynamics prevail.  

It is too early to tell which of the proposed solutions of the measure problem will be finally accepted; for some recent reviews and  proposals see e.g.  \cite{Bousso:2012dk,Vilenkin:2013loa,Guth:2013sya,Linde:2014nna}. It is important that whereas different approaches lead to slightly different constraints on the possible values of the cosmological constant,  these differences do not change the main qualitative conclusion: Many parts of inflationary multiverse are expected to be in a state with a very large absolute value of the cosmological constant; we cannot live there. But there are many exponentially large parts of the multiverse with an extremely small value of the cosmological constant, compatible with our existence.   That is all that we can say with reasonable certainty, and this is already quite sufficient to make the cosmological problem if not completely and unambiguously solved then at least significantly ameliorated. 

Similarly, there is a strong correlation between life as we know it and the values of the electron charge and its  mass, as well as with the ratio between the mass of the proton and the mass of the neutron, and with the number of non-compactified dimensions. Therefore some of the physical parameters describing our part of the world may be environmental, specific to the part of the multiverse where we can live. Meanwhile a discovery of many deep symmetries in the theory of elementary particles suggests that relations between many other parameters may be fundamental rather than environmental. This relates our discussion to    the famous statement made by Einstein in this ``Autobiographical Notes'' \cite{Einstein}:
\begin{quotation}
{\it I would like to state a theorem which at present can not be based upon anything more than a faith in the simplicity, i.e., intelligibility, of nature: There are no arbitrary constants..., nature is so constituted that it is possible logically to lay down such strongly determined laws that within these laws only rationally completely determined constants occur (not constants, therefore, whose numerical value could be changed without destroying the theory)}.
\end{quotation}

The theory of inflationary multiverse does not challenge this philosophical attitude, which is deeply rooted in the basic principles of science. But what we have found is that the total number of truly fundamental constants may be much smaller than what one could expect by studying the observable part of the universe. Sometimes what we perceive as a fundamental constant may be just an environmental parameter, which seems constant because of the effects of inflation, but which may take  entirely different values in other parts of the world. 

An often expressed concern about this theory is that we may not see different parts of the multiverse any time soon, and until this happens there will be no experimental evidence supporting this theory. However, I do believe that we already have  strong experimental evidence in favor of the theory of the multiverse.

In order to explain it, let us take a step back to the time when the inflationary theory was invented. Its main goal was to address many problems which at that time could seem rather metaphysical: Why is our universe so big? Why is it so uniform? Why is it isotropic, why it does not rotate like our galaxy? Why parallel lines do not intersect? It took some time before we got used to the idea that the large size, flatness, isotropy  and uniformity of the universe should not be dismissed as trivial facts of life. Instead of that, they should be considered  as    {\it experimental data} requiring an explanation, which was provided with the invention of inflation. 
 
Similarly, the anomalously small value of the cosmological constant, the extreme smallness of the electron mass, the near coincidence between the proton and neutron masses, as well as the fact that we live in a 4-dimensional space, are {\it experimental data}, and  the only presently available plausible explanation of these and many other surprising experimental results has been found within the general framework of the theory of the multiverse. And, talking about coincidences,  even though possible roles of inflation and string theory in this construction have been conjectured 30 years ago \cite{Linde:1986fd}, the way how different parts of the puzzle started falling into proper places within the context of the string theory landscape was nothing short of miraculous.

That was one of the main reasons why I decided to describe the history of the development of the theory of inflationary multiverse and discuss different pieces of hard to find old papers containing some early and often rather naive formulations of the future theory. Hopefully by looking at it one may either find a possibility to deviate from our original path, or   understand why many of us take the admittedly incomplete progress towards the theory of the inflationary multiverse so seriously.

This work was supported by the SITP,  by the NSF Grant PHY-1316699 and by the Templeton foundation grant `Inflation, the Multiverse, and Holography.'

\section{Appendix A: Eternal chaotic inflation}

Chaotic inflation scenario describes the  scalar field $\phi$ slowly rolling down to the minimum of its potential $V(\phi)$ during the nearly exponential expansion of the universe \cite{Linde:1983gd}. This regime occurs in a broad class of inflationary models with sufficiently flat potentials. It make the universe almost exactly homogeneous on the scale much greater than the typical cosmological scale $H^{-1}$, where  $H = \dot a/a $,  $a(t)$ is a scale factor of the universe. Equation of motion for a homogeneous canonically normalized scalar field in the slow-roll inflationary regime is
\begin{equation}\label{1y}
3H\dot\phi = -V' \ .
\end{equation}
During inflation
\be\label{1z}
H^{2} = V(\phi)/3 \ .
\ee
The universe in this regime expands nearly exponentially, $a(t) \approx e^{H(t) t}$, where $H$ does not change much during the cosmological time $H^{{-1}}$.
Equations (\ref{1y}), (\ref{1z})  imply that during the Hubble time $\delta t = H^{{-1}}$ the field $\phi $ decreases by
\be
\Delta\phi = - V'/V \ .
\ee
This equation suggests that during inflation the field $\phi$ always rolls down towards the minimum of the potential. That is why the possibility that the field may eternally jump against the flow and climb higher and higher  \cite{Linde:1986fc,Linde:1986fd} was so counter-intuitive.

In order to describe this process, one should take unto account that 
space is filled with quantum fluctuations of
all types of physical fields. These fluctuations can be
considered as waves of physical fields with all possible
wavelengths, moving in all possible directions.  
 The wavelengths of all vacuum
fluctuations of the scalar field $\phi$ grow exponentially
in the expanding universe. When the wavelength of any
particular fluctuation becomes greater than $H^{-1}$, this
fluctuation stops oscillating, and its amplitude freezes at
some nonzero value $\delta\phi (x)$ because of the large
friction term $3H\dot{\phi}$ in the equation of motion of the field
$\phi$. The amplitude of this fluctuation then remains
almost unchanged for a very long time, whereas its
wavelength grows exponentially. 

The appearance of
such a frozen fluctuation is equivalent to the appearance of
a classical field $\delta\phi (x)$ that look homogeneous on the scale of the horizon $H^{{-1}}$.
The average amplitude of
such perturbations generated during a time interval $H^{-1}$
is given by \cite{Linde:2005ht}
\begin{equation}\label{E23}
|\delta\phi(x)| \approx \frac{H}{2\pi}\ ,
\end{equation}
which is equal to the Hawking temperature in de Sitter space with the Hubble constant $H$. Each time interval $H^{{-1}}$, new fluctuations of this magnitude and wavelength $H^{{-1}}$ appear and freeze on top of the previously frozen fluctuations.

Thus the evolution of the classical scalar field $\phi$ becomes similar to the Brownian motion: In average, each time $H^{-1}$ the field $\phi$ slides down by $\Delta\phi = - V'/V$. However,  in each part of the universe of a size $H^{{-1}}$ (the size of the horizon) the field may additionally jump either up or down by $\delta\phi(x) \sim  \frac{H}{2\pi} \sim {\sqrt{V}\over 2\sqrt 3\pi}$. The ratio 
\be
{\delta\phi(x)\over \Delta \phi} \sim  {V^{3/2}\over 2\sqrt 3 \pi V'}
\ee
is proportional to the amplitude of {\it post-inflationary}\, scalar perturbations of the metric produced at that time  \cite{Goncharov:1987ir,Creminelli:2008es}. This ratio is very small during the last 60 e-foldings of inflation, but it may be very large in the very early universe.

For example, in the simplest model $V(\phi) = \frac{m^2 \phi^2}{2}$ one has
\be\label{sc}
{\delta\phi(x)\over \Delta \phi} \sim  {m\phi^{2}\over 4\sqrt{6} \pi } \ .
\ee
This means that for $\phi \gg m^{-1/2}$ one has $\delta\phi(x) \gg \Delta \phi$, so the amplitude of jumps up and down is much greater than the overall decrease of the field. Each time $H^{{-1}}$ the size of the universe grows $e$ times, and its volume grows $e^{3} \sim 20$ times. For $\delta\phi(x) \gg \Delta \phi$, this means that the volume of the parts of the universe where the value of the field $\phi$ becomes {\it greater}\, than it was before grows 10 times within the  time interval $H^{{-1}}$. And during the next time interval $H^{{-1}}$, the volume of the parts of the universe with a growing field $\phi$ will grow 10 times again. This leads to eternal  inflation  in the chaotic inflation scenario  \cite{Linde:1986fc,Linde:1986fd}.

Since the scalar perturbations of metric in a post-inflationary universe are proportional to ${\delta\phi(x)\over \Delta \phi}$ (\ref{sc}),  one could wonder whether our considerations concerning eternal inflation are reliable.   However,  scalar perturbations (\ref{sc}) become large only {\it after} inflation. Meanwhile during inflation, the perturbations of energy density during inflation are  very small, of the order $V^{2} \ll V$ for sub-Planckian values of the potential  $V \ll1$  \cite{Goncharov:1987ir,Linde:2005ht}. For example for $m \sim 10^{-5}$, eternal inflation becomes possible at $V(\phi) >10^{{-5}}$, i.e. 5 orders of magnitude  below the Planck density.  
The large amplitude of the scalar perturbations of metric in a post-inflationary universe is a direct consequence of eternal inflation: A sufficiently large part of the universe, which corresponds to inflation starting from $\phi \gg m^{-1/2}$, there will be parts of the universe where inflation is over, and other parts, where eternal inflation still continues. This simply means that the global structure of the universe cannot be described by the FRW metric: The universe becomes a fractal, see Fig. 1.

The conditions for eternal inflation in simplest inflationary models  such as $V(\phi) = \frac{m^2 \phi^2}{2}$ are satisfied all the way up to the Planck density. Close to the Planck density, the Hubble constant could be as large as the Planck mass, and inflationary fluctuations of all fields become maximally strong, experiencing Planck size jumps. This should allow these fields to easily jump from one vacuum state of the theory to another, which could lead not only to modifications in the low-energy laws of physics, but even to jumps between different types of  compactification of space time,  including transitions with different number of non-compactified dimensions \cite{Linde:1982gg,Linde:1986fc,Linde:1986fd,Linde:1988yp,BlancoPillado:2009mi}.

\section{Appendix B: A lower bound on the transition rate between different vacua and string theory landscape}

Consider de Sitter vacuum 
state $\varphi_0$ 
corresponding to the local minimum of the potential with $V_0>0$. Suppose that there is another vacuum $\varphi_{1}$  separated from $\varphi_0$ by a barrier  with $V(\phi) > 0$. What can we say about the possibility of tunneling between these vacua?




To describe tunneling from a local minimum at $\varphi = \varphi_0$ following Coleman and De Luccia   \cite{Coleman:1980aw} one should consider an $O(4)$-invariant Euclidean spacetime with the metric
\begin{equation}\label{metric2}
ds^2 =d\tau^2 +b^2(\tau)(d \psi^2+ \sin^2 \psi \, d \Omega_2^2) \ .
\end{equation}
The scalar field $\varphi$ and the Euclidean scale factor (three-sphere radius) $b(\tau)$ obey the
equations of motion
\begin{equation}\label{equations}
\varphi''+3{b'\over b}\varphi'=V_{,\varphi},~~~~~ b''= -{b\over 3}  (\varphi'^2 +V) \ ,
 \end{equation}
where primes denote derivatives with respect to $\tau$. (We use the system of units $M_p = 1$.)


Coleman-De Luccia  instantons   describe the field $\varphi(\tau)$ beginning in a vicinity of the false vacuum
$\varphi_0$ at $\tau =0$,  and reaching some constant value $\varphi_{f}> \varphi_1$ at $\tau = \tau_{f}$, where $b(\tau_f)= 0$. 
According to \cite{Coleman:1980aw}, the tunneling probability is given by
\begin{equation}\label{prob}
P(\varphi) = e^{-S(\varphi)+S_0},
\end{equation}
where $S(\varphi)$ is the Euclidean action  for the 
tunneling trajectory $(\varphi(\tau),\, b(\tau))$, and   $S_0=S(\varphi_0)$ is the Euclidean action for the initial configuration $\varphi = \varphi_0$.    $S(\varphi)$ in Eq. (\ref{prob}) for the tunneling probability is the integral over the whole instanton solution, rather that the integral over its half  providing the tunneling amplitude.

The tunneling action is given by
\begin{equation}\label{fullaction}
S(\varphi)= \int d^4x
\sqrt{g}\left(-{1\over 2}R+{1\over 2} (\partial \varphi)^2 +V(\varphi)\right).
\end{equation}
In $d=4$ the trace of the Einstein  equation is
$R=(\partial \varphi)^2+4V(\varphi)$. Therefore the total action can be represented by an integral of $V(\varphi)$:
\begin{equation}\label{dsaction}
S(\varphi) = -\int d^4x
\sqrt{g} V(\varphi) = - {2\pi^2\int_0^{\tau_f} d\tau\, b^3(\tau)\, V(\varphi(\tau))}\ .
\end{equation}

The Euclidean action calculated for the false vacuum de Sitter solution $\varphi=\varphi_0$ is given by
\begin{equation}\label{action}
S_0 = - {24\pi^2\over V_0} <0\ .
\end{equation}

This action for de Sitter space $S_0$ has a simple sign-reversal relation to the entropy of de Sitter space ${\bf S_0}$ \cite{Gibbons:1976ue}:
\begin{equation}\label{action2}
{\bf S_0} = - S_0 =+ {24\pi^2\over V_0}\ .
\end{equation}

Therefore the typical time of the tunneling from the de Sitter  vacuum  $\varphi_{0}$ to the de Sitter vacuum $\varphi_{1}$ given by $t_{\rm tunn} \sim P^{-1}(\varphi)$ can be represented in the following way:
\begin{equation}\label{decaytime}
t_{\rm tunn} = e^{S(\varphi)+\bf S_0} = e^{24\pi^2\over V_0} \ e^{S(\varphi)}\ .
\end{equation}
 Eq. (\ref{dsaction}) implies  that for the tunneling through the barrier with $V(\varphi)>0$   the action $S(\varphi)$ is always negative, $S(\varphi) < 0$. This means that {\it the typical time of the tunneling from the de Sitter  vacuum  $\varphi_{0}$ to the de Sitter vacuum $\varphi_{1}$  is always finite and  smaller than  $e^{24\pi^2/ V_0} $}. 
This is very important because it allows tunneling between all de Sitter vacua separated by barriers with $V > 0$, 
even if the barriers are extremely high \cite{Kachru:2003aw}. In other words, all such vacua become interlinked; one may start in one of them and then 

Let us now assume that the lifetime of a de Sitter vacuum is much greater than the typical cosmological time $H^{-1} \sim V_0^{{-1/2}}$. This is a very mild assumption because the probability of tunneling is exponentially suppressed. Then the decay of the de Sitter state never completes in the whole universe. Indeed, the tunneling is a local process which  occurs due to formation of bubbles of the new vacuum. During the exponentially large time required for the tunneling, the non-decayed part of the universe, outside of the bubbles of the new vacuum, continues expanding exponentially, which leads to eternal inflation. This is the same reason why inflation in the old inflationary scenario never ends \cite{Guth:1982pn}, which was a major problem fully resolved only with the invention of the slow-roll chaotic inflation \cite{Linde:1983gd}. In the string theory landscape \cite{Susskind:2003kw}, the slow-roll  inflation may begin {\it after} the tunneling, which is the standard way to solve the the graceful ending problem of the old inflation. 

If the decay of some de Sitter states never completes because of the exponential expansion, the transition from these states to all other vacua of string theory becomes possible. As a result, the universe becomes a multiverse consisting of exponentially many exponentially large parts with different properties.  A more detailed discussion of the tunneling process, including the transitions between different vacua in the situations where the Coleman-De Luccia instantons do not exist, can be  found   \cite{Kachru:2003aw}.

Finally, one should note that the transitions between different vacua is not the only way to implement the string landscape scenario. Even if for some particular reason the transitions between some specific string theory vacua are forbidden (e.g. if they are separated by some exotic barriers with $V<0$), each of these vacua is still a part of a more general inflationary multiverse scenario based on quantum cosmology. Indeed, each of these vacua corresponds to a separate branch of the wave function of the universe, exactly in the same sense as separate universes with different constant values of fluxes $F_{{\mu\nu\lambda\sigma}}$ in the first anthropic solution of the cosmological constant problem, which was proposed long ago in my  review of inflationary cosmology in Reports on Progress in Physics \cite{Linde:1984ir}.

\newpage

\centerline{\LARGE{\bf Addendum: Excerpts from some early papers}}

\

\ 

\centerline{\Large{\it {1. On eternal inflation, anthropic principle, and multiverse:}}}

\

\

{\small

\begin{flushright}
    \hfill{July 1982 }
\end{flushright}

\begin{center}
    { \Large{\bf Nonsingular Regenerating Inflationary Universe}}\footnote{This preprint was published during the Nuffield Symposium  in Cambridge, July 1982. It contains one of the first descriptions of eternal inflation in the new inflation scenario. (I introduced the name `eternal inflation' several years later, in 1986.) Unfortunately, the idea to avoid the initial singularity problem due to eternal inflation, which was introduced in this preprint, did not work in the way I expected at that time. However, the picture of an eternally inflating universe divided into many exponentially large parts with dramatically different properties, and the possibility to use the anthropic principle in this context, which was first introduced in this paper, became a definitive part of the theory of inflationary multiverse. A more detailed description of this theory was given in my contribution to the Proceedings of the Nuffield Symposium.} 
    
    \vspace{10pt}
    
  {\large  {\bf  Andrei Linde }}
    
    \vspace{10pt}
    
    {\it University of Cambridge, DAMTP, Silver St., Cambridge CB3 9EW, England}\\
        \vspace{3pt}
    
    Permanent address: {\it Lebedev Physical Institute, Moscow 117924, Leninsky prospect 53, USSR}
    
  \end{center}

\begin{abstract}
    
A new version of the inflationary universe scenario is suggested which provides a possible solution of the cosmological singularity
problem.

\end{abstract}

\vspace{10pt}

\baselineskip 5.5 mm

There is a large interest now in the inflationary universe scenario [1 - 12], which may provide us with a solution of many cosmological problems
such as the horizon, flatness and primordial monopole problems [1], homogeneity, isotropy and domain wall problems [2], and the problem of the
origin of fluctuations which are necessary for the galaxy formation [11, 6, 12]. A lot of work is to be done before any final version of this scenario will
be elaborated, but the results obtained up to now [1 - 12] seem rather
encouraging. However, so far it has not been quite clear whether the inflationary
universe scenario, which may help us to solve so many cosmological problems,
can help us also to solve the most important cosmological problem, the
problem of the cosmological singularity. It is the aim of the present
paper to suggest a possible solution of this problem in the context of
the inflationary universe scenario.

In order to make things as simple as possible let us first recall some
features of the inflationary universe scenario. In the first version of
this scenario [1] it was assumed that the phase transition from the metastable supercooled vacuum state $\phi= 0$ to the true vacuum state $\phi = \phi_{0}$
proceeds by formation and expansion of bubbles of the field $\phi$ inside the
metastable phase $\phi = 0$. It was assumed that the field $\phi$ inside the bubble
almost instantaneously grows up to its equilibrium value $\phi = \phi_{0}$, and the
velocity of the bubble walls rapidly approaches the velocity of light
$c = 1$. Thermalization in this scenario could occur only due to bubble
wall collisions. However, Guth and Weinberg have shown [13] that the
probability of the collision of the bubble formed at some time t with the
bubbles created earlier asymptotically (at large t) approaches the
t-independent value $N = {80\pi\epsilon\over 9}$, where $\epsilon = {\lambda\over H^{4}}$, $\lambda$ is the bubble formation
probability per unit time per unit physical volume, $H$ is the Hubble
constant. Therefore if the rate of the bubble production is sufficiently
small, $\epsilon \ll {9\over 80\pi}$, the newly formed bubbles never collide with the bubbles
formed earlier, and the phase transition never completes (non-percolation
of bubbles [13]). We come to a very unusual and interesting picture of
the boiling and self-reproducing (regenerating) exponentially expanding
universe. Unfortunately, however, such a universe is not the best place to
live in, since thermalization in this universe does not occur, and both
inside and outside the bubbles there is no matter at all [13].

In the simplest versions of the new inflationary universe scenario
[2, 4, 10] the phase transition also proceeds by formation of bubbles
of the field $\phi$. However in this scenario the field $\phi$  inside the bubble
at the moment of its formation is much smaller that $\phi_{0}$. After the bubble
formation the radius of the bubble exponentially increases, whereas the
field $\phi$ grows very slowly. Therefore one bubble covers all the observable
part of the universe before the field $\phi$ inside the bubble grows up to
$\phi \sim \phi_{0}$. Thermalization in this scenario occurs not due to the bubble wall
collisions, but due to the creation of particles by the almost homogeneous
field $\phi$ convergently oscillating near its equilibrium value $\phi \sim \phi_{0}$ [2, 4, 7].
Let us note, however, that the bubble wall velocity in this scenario (not
to be confused with the rate of growth of the bubble radius) is smaller
than that in the old inflationary universe scenario, and approaches the velocity
of light only after the field $\phi$ inside the bubble becomes sufficiently
large. Therefore in the new inflationary universe scenario it is even more
difficult for the bubbles to collide than in the old scenario, and from
the results of ref. [13] it follows that at least for $\epsilon \ll {9\over 80\pi}$ the phase
transition is never completed in the whole universe. However, in the new
inflationary universe scenario this fact is not dangerous at all, since
thermalization occurs in each bubble separately,\footnote{The possibility of such a realization of the new inflationary universe
scenario with singularity at the initial stage of the universe evolution
and with the regenerative regime at the inflation stage was first noticed
in [8].} and each bubble can be
considered as a mini-universe which has no possibility of any causal contact
(e.g. by collision) with other mini-universes.

 This statement will become more precise if we call the ``bubble" the
region of space-time which can be somewhat affected by the formation of the
bubble of the field $\phi$. The boundary of this region moves with velocity exactly
equal to that of the speed of light. Such a region has a size slightly
greater than the size of the bubble of the field $\phi$. With this new definition
of the ``bubble" all analysis of the bubble collisions performed by Guth and
Weinberg [13] remains unaltered. This means that for $\epsilon \ll {9\over 80\pi}$ most
of the physical volume of the universe will never know about the existence of
any bubbles, and therefore the universe as a whole (and, in particular, the
space between non~percolating bubbles) will remain exponentially expanding
as if there were no bubbles at all [13]. 

Now we come to the most important part of our paper. Usually it is
assumed that the universe initially was hot and singular, and only after some
cooling of the universe could start the phase transition discussed above.
However in the new inflationary universe scenario this assumption is not
obligatory at all. As we have seen already, if the rate of the bubble formation
is sufficiently small, the phase transition is never completed in the whole
universe, and at $t\to \infty$ we have an exponentially expanding boiling self-reproducing
universe, which is cold (has some small constant Hawking temperature $T = {H\over 2\pi}$)
everywhere outside some small fraction of the physical volume of the universe
occupied by the bubbles. But since the universe can live in this self-
reproducing stationary state forever, we may relax now our usual assumption
that the universe initially was hot, and assume that the universe as a whole
always was in this regenerating state. This assumption, which is in complete
agreement with our previous analysis, means that our universe as a whole
never was singular (e.g. the curvature scalar $R = 12 H^{2} = const$ everywhere
outside the bubbles, and is even smaller inside the bubbles). There was no
beginning and there will be no end to the universe's evolution. However each
mini-universe (each bubble) has its beginning in time (the time at which
this bubble was formed without any coming through singularity), and its
evolution after thermalization can be described by the usual hot universe
theory.

Thus we see that the new inflationary universe scenario may provide us
with a solution of the cosmological singularity problem. 


Several comments are in order.

1. The resolution of the singularity problem in the context of the scenario,
in which many mini-universes are created from the metastable vacuum state,
resembles some attempts to solve the singularity problem by assuming creation
of the universe from ``nothing" [14] or from some other universe [15]. However
in distinction to our scenario, in the latter approaches to the singularity
problem it is assumed that there was some moment before which our physical
space and time have not existed. From our point of view such a resolution
of the singularity problem would be somewhat incomplete, since the most
important part of this problem is just a question of how it could be that
sometimes(?) our space and time have not existed. Of course, one could argue
that near the cosmological singularity our usual concepts of space and time
do not work, and the resolution of the singularity problem might be found
after the proper development of the quantum gravity theory. It seems to us,
however, that it would be not too bad to have a possible solution of the
singularity problem which would not require any crucial changes of the usual
concepts of space and time in the very early universe.

{  2. It is sometimes assumed that there might be many different universes,
and we live in just one of them, which is sufficiently suitable for the
existence of life and human consciousness. This assumption is one of the
possible motivations of the Anthropic Principle in cosmology (for a
recent discussion of some relevant issues see e.g. [16, 17]) However,
it was not quite clear whether God actually could play the game of the universe
creation many times before the final success, or maybe one could try to
make some sense out of this many-universe hypothesis e.g. in the context of
quantum gravity. It may be of some interest therefore that in the scenario
suggested above the universe contains an infinite number of mini-universes (bubbles) of different size, and in each of these universes the masses of particles, coupling constants etc. may be different due to the possibility of different
symmetry breaking patterns inside different bubbles. This may give us a
possible basis for some kind of Weak Anthropic Principle: There is an
infinite number of causally unconnected mini-universes inside our universe,
and life exists only in sufficiently suitable ones.}

3. The nonsingular regenerating inflationary universe scenario suggested
above can actually be realized only in theories in which bubble
formation is sufficiently strongly suppressed (small $\epsilon$) On the other hand,
in order to have a large inflation the curvature of the effective potential
$V(\phi)$ near $\phi = 0$ should not be too large [2-4, 6, 10]. Presumably both
these conditions can be satisfied in a theory with a small positive mass
squared of the field $\phi$, $m^{2}(\phi = 0) \ll H^{2}$, and with a sufficiently small scalar
coupling constant $\lambda$ [2 6, 10].

It is worth noting that the nonsingular inflationary universe scenario
can be implemented not only in the context of grand unified theories, but
in the context of quantum gravity as well. For example, a sufficiently
large inflation of the universe can be achieved in the Starobinsky model [18].
However, according to this model the universe initially was in the unstable
vacuum state. The lifetime of the universe in such a vacuum state is
finite [11] (as distinct from the lifetime of the universe in the regenerative
state considered in the present paper). This means that such a vacuum state
could not exist at $t \to \infty$,\footnote{A Similar comment can be applied also to a recent paper by Atkatz and
Pagels [19], in which it is assumed that the lifetime of the universe
in the initially nonsingular state is finite.}
 and therefore for realization of the
Starobinsky scenario it is necessary for the universe to be created in
the unstable vacuum state either from ``nothing" [14] or from some other
universe [15]. Recently a modification of this model was suggested, in which
the vacuum state initially was metastable rather than unstable [20]. The
theory of the vacuum decay in this model almost coincides with that in the
inflationary universe scenario. One may expect therefore that the nonsingular
regenerating inflationary universe scenario can be implemented also in the
context of the modified Starobinsky model mentioned above.

We see therefore that there exist several different versions of the
new inflationary universe scenario, which may help us to solve not only the
horizon, flatness, homogeneity and isotropy problems, but also the cosmological singularity problem. Of course, our present understanding of
the phase transitions in the exponentially expanding universe is still
far from being complete, and a more detailed analysis of this question
 is needed in order to verify whether the nonsingular inflationary
universe scenario can be implemented in the context of realistic theories
of elementary particles. In any case, however, the very possibility of
obtaining a solution to all the most important cosmological problems in the
context of a unique rather simple scenario seems very interesting and
clearly deserves further investigation.

It is a pleasure to express my deep gratitude to the participants of
the Nuffield Workshop on the Very Early Universe in Cambridge, and
especially to A.H. Guth, S.W. Hawking and P.J. Steinhardt for many
enlightening and stimulating discussions. I am also thankful to S.W. Hawking
and to Cambridge University for their hospitality during the Nuffield
Workshop in Cambridge.

\

{\bf \Large \noindent References}

\begin{enumerate}

\item A.H. Guth, Phys. Rev. D23 (1981) 347.
\item  A.D. Linde, Phys.Lett. 108B (1982) 389;\\
~~~~~~~  A.D. Linde, Coleman-Weinberg theory and a new inflationary universe
scenario, Lebedev Phys. Inst. preprint No. 88 (March 1982), to be
published in Phys. Left. B;\\
A.D. Linde, Temperature dependence of coupling constants and the phase
transition in the Coleman-Weinberg theory, Lebedev Phys. Inst. preprint
No. 121 (May 1982), submitted to Phys. Lett. B;\\
A.D. Linde, Scalar field fluctuations in expanding universe and the
new inflationary universe scenario, submitted to Phys. Lett. B.
\item S.W. Hawking and I.G. Moss, Phys. Lett. 110B (1982) 35.

\item A. Albrecht and P.J. Steinhardt, Phys. Rev. Lett. 47 (1982) 1220; A. Albrecht, P.J. Steinhardt, M. Turner and F. Wilczek, Phys. Rev. Lett. 47 (1982) 1437.
\item A. Vilenkin, Tufts University preprint TUTP-82-5 (1982);
 L. Ford and A. Vilenkin, Tufts University preprint TUTP-82-7 (1982).
\item A.A. Starobinsky, a talk at the Nuffield Workshop on the Very Early Universe, Cambridge (June 1982), to be published.
\item A.D. Dolgov and A.D. Linde, Baryon asymmetry in inflationary universe, preprint ITEP-78 (May 1982), submitted to Phys. Lett. B.
\item P.J. Steinhardt, a talk at the Nuffield Workshop on the Very Early Universe, Cambridge (June 1982), to be published.
\item I.G. Moss, a talk at the Nuffield Workshop on the Very Early Universe, Cambridge (June 1982), to be published.
\item A.S. Goncharov and A.D. Linde, Bubble formation in the inflationary universe (in preparation).
\item	G.V. Chlbisov and V.F. Mukhanov, Pisma ZhETF 33 (1981) 549;\\
G.V. Chibisov and V.F. Mukhanov, Lebedev Phys. Inst. preprint No. 198 (1981), to be published in ZhETF.
\item	S.W. Hawking, The development of irregularities in a single bubble inflationary universe, DAMTP preprint (June 1982);\\
J.M. Bardeen, P.T. Steinhardt and M.S. Turner, to be published;\\
 A.H. Guth, a talk at the Nuffield Workshop on the Very Early Universe,
Cambridge (June 1982), to be published.
\item A.H. Guth and E.T. Weinberg, MIT preprint CTP No. 950 (1982), to be published in Nucl. Phys. B.
\item E.P. Tryon, Nature 246 (1973) 396;\\
Ya.B. Zeldovich, Pisma Astron. Zh. 7 (1981) 579;\\
L.P. Grishchuk and Ya.B. Zeldovich, in: Quantum structure of space and time (Cambridge University Press, ed. M. Duff and C. Isham, 1982); \\
A. Vilenkin, Tufts University preprint TUTF-82-8.
\item R. Brout, F. Englert and E. Gunzig, Ann. Phys. 115 (1978) 78.
\item S.W. Hawking, The cosmological constant and the Weak Anthropic Principle, DAMTP preprint (1982).
\item I.L. Rosental, Soviet Physics: Uspekhi 131 (1980) 239.
\item A.A. Starobinsky, Phys. Lett. 91B (1980) 99.
\item D. Atkatz and H. Pagels, Phys. Rev. D25 (1982) 2065.
\item V.Ts. Gurovich, Pisma Astron. Zh., to be published.

\end{enumerate}

\newpage 

\centerline {\LARGE \bf  {The New Inflationary Universe Scenario}}

\

{ \bf from:  \ A.D. Linde, In: {\it   The Very Early Universe}, ed. G.W. Gibbons, S.W. Hawking and S.Siklos,
Cambridge University Press (1983),   pp. 205-249 (Proceedings of the Nuffield Symposium, June-July 1982)}\footnote{This is a part of my contribution of the Nuffield Symposium  in Cambridge, July 1982. It contains a more detailed discussion of inflationary multiverse and anthropic principle.} 
 
\
 
{\bf Appendix B. INFLATIONARY UNIVERSE SCENARIO AND THE ANTHROPIC PRINCIPLE}

As we have mentioned in the Introduction, there exists a possible solution of the flatness. homogeneity, isotropy and baryon asymmetry problems based on the Anthropic Principle: It can be argued that in a curved, inhomogeneous and anisotropic universe without any baryon asymmetry no life would exist, and therefore nobody would ask any questions about homogeneity, isotropy, etc. This argument is very witty but is not quite 
convincing, since it cannot explain why the spectrum of inhomogeneities is 
almost scale-independent, why the universe is almost exactly isotropic and 
why the numerical value of $(n_{B}-n_{\bar B})/n_{\gamma}$ is $0(10^{-8})$. 

In some cases, however, the results obtained by the use of the 
Anthropic Principle may be very informative. For example. one can obtain 
an ``explanation" why our space is four-dimensional. Indeed, in the theories of electromagnetic and gravitational interactions in d-dimensional 
space-time at $d > 4$ any bounded systems such as atoms or planetary systems would be impossible, whereas at $d < 4$ free particles cannot 
exist. This indicates that the best conditions for the existence of life 
(or, at least, of our kind of life, based essentially on electromagnetic 
and gravitational interactions) can be realized just as in the four-dimensional space-time (Ehrenfest 1917). Similar considerations based on the 
study of conditions necessary for the existence of atoms, stars, galaxies,
etc. lead to stringent constraints on the value of $\alpha = {e^{2}\over 4\pi}$, on the value 
of the gravitational constant G and on the masses of elementary particles 
(Carr \& Rees 1979; Rosental 1980). The Anthropic Principle may help us to 
understand why the cosmological constant at present is zero (Hawking 
1982a), why the elementary particle theory has the unbroken symmetry 
$SU(3) \times U(1)$ etc. 
It could be argued, however, that our universe is unique. and 
it is meaningless to ask whether life can exist in space with $d \not = 4$ and 
with $\alpha \not = {1\over 137}$. One possible answer to this objection is that there may 
exist many disconnected universes (see Appendix A), and we live in just 
one of them, which is sufficiently suitable for the existence of intelligent life. Another possibility is related to the oscillating universe 
scenario discussed in Appendix A, since in this scenario the universe is 
created anew in each new cycle. However the simplest way to justify the 
Anthropic Principle (at least partially) can be found in the context of 
the new inflationary universe scenario. 

Indeed, the phase transition in the new inflationary universe 
scenario occurs independently in each bubble, and after the phase transition the distance between the centers of the bubbles exceeds the size of 
the horizon $\sim ct$, where $t$ is the age of the universe. Therefore the 
physical processes inside a bubble remain unaffected by the physical processes inside any other bubble even after the phase transition. In this 
sense any bubble behaves as a mini-universe almost completely isolated 
from all other mini-universes. Let us assume e.g. that the universe is 
open. After the phase transition this universe is divided into an infinite 
number of mini-universes, in each of which the phase transition occurs independently. This means that there will be infinitely many mini-universes 
with all possible symmetry breaking patterns. In particular, in the $SU(5)$
Coleman-Weinberg theory the phase transition may proceed with a comparable 
probability either to the phase $SU(3) \times SU(2) \times U(1)$ or to the phase 
$SU(4) \times U(1)$. Therefore after the phase transition the universe becomes 
divided into infinitely many mini-universes in the phase $SU(3) \times SU(2) \times U(1)$
and infinitely many mini-universes in the phase $SU(4) \times U(1)$. This considerably simplifies the answer to the question why we are now in the phase 
$SU(3) \times U(1)$ but not in some other stable or metastable phase. corresponding to a local minimum of the effective potential. Indeed. the phase 
$SU(3) \times SU(2) \times U(1)$ eventually evolves into the phase $SU(3)\times U(1)$. Therefore this phase (if it is sufficiently stable) should exist inside infinitely many 
mini-universes, and all other ``desert islands" (in which other 
kinds of life may exist) are of no importance for us. 

There is some interest now in the theories in which our space 
originally has more than four dimensions. but extra dimensions are spontaneously compactified (see e.g. Cremmer \& Scherk 1976; Witten 1981a). However, it is not very simple to explain why just four dimensions remained 
uncompactified (Schwarz 1982). The new inflationary universe scenario 
makes it more easy to answer this question. In the context of this scenario it would be sufficient that the compactification to the space $d = 4$ 
is {\it possible}, but there is no need for the four-dimensional space to be the 
{\it the only} possible space after the compactification. Indeed, if the compactification to the space $d = 4$ is possible, there will be infinitely many 
mini-universes with $d = 4$ in which intelligent life can exist. 

Thus we see that the new inflationary universe scenario may 
serve as a basis for a kind of a Weak Anthropic Principle: All possible 
phases which may appear after the phase transition and which are sufficiently stable should exist in some of the mini-universes. and then 
(hopefully) we are free to choose the best mini-universe to live in. As 
was claimed by Guth (1982), the inflationary universe is the only example 
of a free lunch (all matter in this scenario is created from the unstable 
vacuum). Now we can add that the inflationary universe is the only lunch 
at which all possible dishes are available.

\

REFERENCES:

\begin{enumerate}
\item Ehrenfest, P. (1917). Proc. Amsterdam Acad. {\bf 20}, 200.
\item Carr, B.J and  Rees, M.J. (1979). Nature {\bf 278} 605. 
\item Rozental' I. L. (1980). Usp. Fiz. Nauk {\bf 131}, 239  [Sov. Phys. Usp. {\bf 23}, 296]
\item Cremmer, E. and  Scherk, J. (1976). {\it Spontaneous compactification of space in an Einstein-Yang-Mills-Higgs model}. Nucl. Phys. {\bf B108}, 409-416.
\item Witten, E.  (1981a). {\it Search for a realistic Kaluza-Klein theory.} Nucl. Phys. {\bf B186}, 412-428.
\item Schwarz, J.H. (1982). {\it Superstring theory.} Caltech preprint CALT-68-9111.
\item Guth, A.H. (1982). {\it $10^{-35}$ seconds after the Big Bang.} MIT preprint CTP \# 991.
\end{enumerate}

\newpage

\centerline{\Large{\it 2. On anthropic principle and cosmological constant}}

\

\
  
\centerline {\LARGE \bf  {The Inflationary Universe}}

\

\centerline{\bf  from: A.~D.~Linde,  Rept.\ Prog.\ Phys.\  {\bf 47}, 925 (1984).}

\

\centerline{\bf the last paragraph in  Conclusions\footnote{A possible solution of the cosmological constant problem based on inflation, quantum cosmology, and anthropic considerations.}}

\

Another possible way of solving the cosmological constant problem is related to quantum cosmology. The vacuum energy density may depend on the topology of the compactified part of space (Sakharov 1984) and on some classical fields of the type of the antisymmetric tensor field $A_{\mu\nu\lambda}$ (Ogievetsky and Sokatchev 1980, Duff and van
Nieuwenhuizen 1980, Aurelia {\it et al} 1980). This field, just as the scalar field $\phi$, appears simultaneously with quantum creation of the universe (Hawking 1984a, b). However, contrary to the scalar field $\phi$, the field strength $F_{{\mu\nu\lambda\sigma}}$ of  the field $A_{\mu\nu\lambda}$, which gives the contribution $V(F)$ to the vacuum energy $V(\phi, F)= V(\phi)
+V(F)$, remains constant during the subsequent classical evolution of the universe (Ogievetsky and Sokatchev
1980, Duff and van Nieuwenhuizen 1980, Aurelia {\it et al} 1980). As follows from equation (13.7), the universe is created most probably in a state with $V(\phi, F) \sim M_{p}^{4}$: (Linde 1984c,d, Starobinsky 1984b). However, this does not impose any constraints on the value of $V(F) =V(\phi, F)- V(\phi)
$ since the value of $V(\phi)$ at the initial stages of inflation can be arbitrarily large (see Sect. 13). Therefore, {\it after} symmetry breaking any value of vacuum energy density $V(\phi, F)= V(\phi)
+V(F)$ may appear with approximately the same probability. At $|V(\phi, F)| \gg  10^{-29} {\rm g\cdot cm^{-3}}$ life of our type would be impossible. The value $|V(\phi, F)| \lesssim  10^{-29} {\rm g\cdot cm^{-3}}$ {\it a priori} does not seem very probable. The eternally oscillating universe scenario is not of much help here, since the closed universe with $V(\phi, F)>0$ can expand forever, which would break the chain of oscillations (see Sect. 13). What may occur, however, is a multiple quantum production of ÔnewÕ universes from the ÔoldÕ ones. Such a process looks like an infinite chain reaction, which is possible due to the gravitational instability discussed in Sect. 13 (see also a paper by Englert and Nicolai (1983) in which similar ideas were suggested). During this process infinitely many universes can be produced, in some of which $|V(\phi, F)| \lesssim  10^{-29} {\rm g\cdot cm^{-3}}$ and life of our type may exist. This is a possible solution of the cosmological constant problem based on the implementation of the anthropic principle in quantum cosmology.

\

\newpage

\centerline {\Large \bf  {Cosmological Transitions With A Change In Metric Signature}}

\

\centerline {Section 4 from the paper}

\vskip 2mm

\centerline{\bf  A.~D.~Sakharov,  Sov.\ Phys.\ JETP {\bf 60}, 214 (1984)
  [Zh.\ Eksp.\ Teor.\ Fiz.\  {\bf 87}, 375 (1984)].\footnote{A possible solution of the cosmological constant problem based on the multiplicity of different types of compactification.}}

\
 
4. ANTHROPIC PRINCIPLE AND THE COSMOLOGICAL
CONSTANT

The different regions of the $Q$ space may differ in their
discrete and continuous parameters. In the spirit of the anthropic
principle we assume that the observed Universe is
distinguished by a set of values of the parameters favorable
for the development of life and intelligence. In particular, it
is possible that the signature (equal to 1 or another odd number)
is one such parameter.

For a Universe with given signature, as further discrete
parameters we must consider the number of dimensions of
the compactified factor space K and the macrospace
$M = Q - a - K$, which is not necessarily equal to 3. This
possibility, which follows from the compactification hypothesis,
is a natural realization of the idea$^{1}$ that Universes
with different numbers of spatial dimensions M arise; evidently,
Ehrenfest's arguments$^{2}$ for the reason why ``our"
case $M = 3$ is distinguished remain valid.

The topological characteristics of the boundaries of the
P and U regions are also discrete parameters. The discrete
parameters determine the effective Lagrangian of the macrospace.
The continuous parameters are the initial values of the
characteristics of the matter fields and the initial deviations
from symmetry of the transition boundaries. These parameters
together with the discrete parameters determine the
evolution of the Universe.

It is well known that the cosmological constant is zero,
$\Lambda = 0$, or anomalously small and, moreover -and this is
particularly remarkable-not in the internally symmetric
state of the ``false" vacuum but in the state of the ``true"
vacuum with broken symmetries. The smallness or vanishing
of $\Lambda$ is one of the main factors that ensures a prolonged
existence of the Universe, sufficient for the development of
life and intelligence. It is therefore natural to invoke the anthropic
principle to solve the problem of the cosmological
constant.

If the small value of the cosmological constant is determined
by ``anthropic selection," then it is due to the discrete
parameters. At the same time, $\Lambda$ is either exactly equal to
zero in some variant, or exceptionally small. In the latter
case, it must be assumed that the number of variants of the
set of discrete parameters is sufficiently large to make the
spectrum of $\Lambda$ values in the neighborhood of the point $\Lambda$ = 0
sufficiently ``dense." This obviously requires a large value of
the number of dimensions K of the compactified space or
(and) the presence in some topological factors of a complicated
topological structure (such as a large number of ``arms").

We note in conclusion that in $P$ space one must consider
an infinite number of $U$ inclusions (for the complete set of
trajectories or even for one trajectory); at the same time, the
parameters of an infinite number of them may be arbitrarily
close to the parameters of the observed Universe. Therefore,
it can be assumed that the number of Universes similar to
ours, in which structures, life, and intelligence are possible,
is infinite. This does not rule out the possibility that life and
intelligence are also possible in an infinite number of very
different Universes that form a finite or infinite number of
classes of ``similar" Universes, including Universes with signature
different from ours.

\begin{enumerate}

\item
  B. J. Carr and M. J. Rees, Nature {\bf 278}, 605 (1979); I. L. Rozental', Usp.
Fiz. Nauk Usp. Fiz. Nauk {\bf 131}, 239  (1980) [Sov. Phys. Usp. {\bf 23}, 296 (1980)]; A. D. Linde,
in: The Very Early Universe (eds. G. Gibbons, S. Hawking, and S. Siklos). Cambridge Univ. Press (1983), p. 205.
\item P. Ehrenfest, Proc. Amsterdam Acad. {\bf 20}, 200 (1917).

\end{enumerate}

\




\

\centerline {\LARGE \bf  {Inflation and quantum Cosmology}}

\

\centerline{\bf from:  A. D. Linde, Print-86-0888, 1 July 1986; published in: {\it 300 Years of Gravitation}\footnote{A possible anthropic solution of the cosmological constant problem in the simplest version of the theory of dark energy.}}

\vskip 3pt

\centerline{\bf  Eds. S. W. Hawking and W. Israel, Cambridge University Press, Cambridge, 1987}

\

In order to illustrate new possibilities which appear in the context of the scenario discussed above, let us consider now a toy model which may explain the small value of the vacuum energy density $\rho_{v}$ (of the cosmological constant), in the observable part of the universe ($|\rho_{v}| \lesssim 10^{-29} {\rm g\cdot cm^{-3}}$). This model is probably unrealistic, but nevertheless it may be rather instructive. The model describes an inflaton field $\phi$, which drives inflation, and a field $\Phi$ with an extremely flat effective potential, $V(\Phi) = \alpha M_{p}^{3}(\Phi-C)$, where $\alpha \lesssim 10^{{-120}}$. It can be shown that during the time interval $t \sim 10^{10}$ years after inflation such a field $\Phi$ remains essentially unchanged due to the very small slope of $V(\Phi)$. However, the Brownian motion of this field during inflation is very rapid, and it divides the universe into an exponentially large number of mini-universes containing {\it all} possible values of the field $\Phi$ for which $|V(\phi) + V(\Phi)| < M_{p}^{4}$. After inflation the vacuum energy $\rho_{v}$ inside these mini-universes is given by $V(\phi, \Phi) =  \alpha M_{p}^{3}(\Phi-C)+ V(\phi_{0}))$, where $V(\phi_{0})$ corresponds to the minimum of $V(\phi)$. This quantity in different mini-universes changes continuously from $-M_{p}^{4}$ to $M_{p}^{4}$, but life of our type is possible only in those mini-universes in which $|\rho_{v}| \lesssim 10^{-29} {\rm g\cdot cm^{-3}}$. Indeed, domains with $V(\phi, \Phi)<0$, $|V(\phi, \Phi)| \gg 10^{-29} {\rm g\cdot cm^{-3}}$ correspond to anti-de Sitter mini-universes with a lifetime $t\ll 10^{10}$ years. The domains with $V(\phi, \Phi)\gg 10^{-29} g\cdot cm^{{-3}}$ remain inflationary for $t\gtrsim 10^{10}$ years, and the present density of matter in such domains is negligibly small. For this reason we see ourselves inside a domain with $|\rho_{v}| \lesssim 10^{-29} {\rm g\cdot cm^{-3}}$.

To make this model realistic it would be desirable to explain why $V(\phi)$ is so flat (though similar potentials sometimes appear in realistic models describing unified theories of elementary particles). In any case, the main idea suggested above may be used in other models as well. In our scenario it is not necessary to insist (as is usually done) that the vacuum energy must disappear in a `true' vacuum state. It is quite sufficient if there exists some relatively stable vacuum-like state with $|\rho_{v}| \lesssim 10^{-29} {\rm g\cdot cm^{-3}}$. This requirement is still very restrictive, but it can be satisfied much more easily than the previous one. For a discussion of a similar approach to the
cosmological constant problem, see also Sakharov (1984) and Linde (1984c).

}

\newpage

\

\centerline{\Large{\it 3. On eternal chaotic inflation and string theory landscape}}

\

\

\centerline {\LARGE \bf  {Eternally Existing Self-reproducing}}

\vskip 8pt

\centerline{\LARGE \bf  Chaotic Inflationary Universe}

\

\centerline{\bf from:  A.~D.~Linde,
Phys.\ Lett.\ B {\bf 175}, 395 (1986).}

\

\

It is worth noting that the process of self-reproduction of the universe occurs not only at the Planck density, but at much smaller densities as well, e.g. at $V(\phi)>\lambda^{1/3} M_{p}^{4} \sim 10^{-4}M_{p}^{4}$ for $\lambda \sim 10^{-12}$. Therefore to prove the very existence of the regime of self-reproduction of inflationary mini-universes in our scenario there is no need to appeal to unknown physical processes at $\rho > M_{p}^{4}$.
On the other hand, it is very important that independently of the origin of the universe in our scenario (either the universe was created as a whole at $t = t_{p}$ or it exists eternally) it now contains an exponentially large (or even infinite) number of mini-universes, and a considerable part of these mini-universes was created when the field $\phi$ was $O(\lambda^{1/3} M_{p})$ and its energy density was $O(M_{p}^{4})$. (Note, that this is true in the chaotic inflation scenario only, in which inflation may occur even at $V(\phi)\sim M_{p}^{4}$.) At such densities, fluctuations of all fields and fluctuations of metric are very large at a typical scale $\sim H^{-1} \sim M_{p}^{{-1}}$. This may lead to the generation of different classical scalar fields $\Phi_{i}$, corresponding to different local minima of $V(\phi, \Phi_{i})$ in different domains of the universe and to processes of compactification or decompactification which occur independently in each of the causally disconnected mini-universes of  initial size $l\gtrsim H^{-1} \sim M_{p}^{{-1}}$. 

As a result, our universe at present should contain an exponentially large number of mini-universes with {\it all}\, possible types of compactification and in {\it all}\, possible (metastable) vacuum states consistent with the existence of the earlier stage of inflation. If our universe would consist of one domain only (as it was believed several years ago), it would be necessary to understand why Nature has chosen just this one type of compactification, just this type of symmetry breaking, etc. At present it seems absolutely improbable that all domains contained in our exponentially large universe are {\it of the same type}. On the contrary, {\it all}\, types of mini-universes in which inflation is possible should be produced during the expansion of the universe, and it is unreasonable to expect that our domain is the only possible one or the best one. 

From this point of view, an enormously large number of possible types of compactification which exist e.g. in the theories of superstrings should be considered not as a difficulty but as a virtue of these theories, since it increases the probability of the existence of mini-universes in which life of our type may appear. The old question why our universe is the only possible one is now replaced by the question in which theories the existence of mini-universes of our type is possible. This question is still very difficult, but it is much easier than the previous one. In our opinion, the modification of the point of view on the global structure of the universe and on our place in the world is one of the most important consequences of the development of the inflationary universe scenario.

\


\end{document}